# ETMA: Efficient Transformer Based Multilevel Attention framework for Multimodal Fake News Detection

Ashima Yadav[1], Shivani Gaba[2], Haneef Khan[3], Ishan Budhiraja[4], Akansha Singh[5], and Krishan Kant Singh[6]

*School of Computer Science Engineering and Technology, Bennett University, Greater Noida, India*[1,2,4,5], *College of Computer Science and Information Technology, Jazan University,* Jazan[3]*, Department of Computer Science Engineering, ASET, Amity University, Noida*[6]

ashimayadavdtu@gmail.com[1], E20SOE822@bennett.edu.in [2], hanees.khan@jazanu.edu.sa[3], ishan.budhiraja@bennett.edu.in[4], akansha.singh1@bennett.edu.in[5], kksingh@amity.edu[6]

**Abstract:** In this new digital era, social media has created a severe impact on the lives of people. In recent times, fake news content on social media has become one of the major challenging problems for society. The dissemination of fabricated and false news articles includes multimodal data in the form of text and images. The previous methods have mainly focused on unimodal analysis. Moreover, for multimodal analysis, researchers fail to keep the unique characteristics corresponding to each modality. This paper aims to overcome these limitations by proposing an Efficient Transformer based Multilevel Attention (ETMA) framework for multimodal fake news detection, which comprises the following components: visual attention-based encoder, textual attention-based encoder, and joint attention-based learning. Each component utilizes the different forms of attention mechanism and uniquely deals with multimodal data to detect fraudulent content. The efficacy of the proposed network is validated by conducting several experiments on four real-world fake news datasets: Twitter, Jruvika Fake News Dataset, Pontes Fake News Dataset, and Risdal Fake News Dataset using multiple evaluation metrics. The results show that the proposed method outperforms the baseline methods on all four datasets. Further, the computation time of the model is also lower than the state-of-the-art methods.

## 1. Introduction

Fake news can be defined as deliberately tampering the information by manipulating images or text to mislead the readers. People create attractive images and falsified text to attract more viewers and engage new readers. This creates a harmful impact as people's thoughts can be manipulated, which can cause serious public concern. The earlier approaches have considered several fusion-based techniques [1] [2] [3] for detecting fake news from multimodal data and paid less attention to the individual characteristics of the modalities [4].

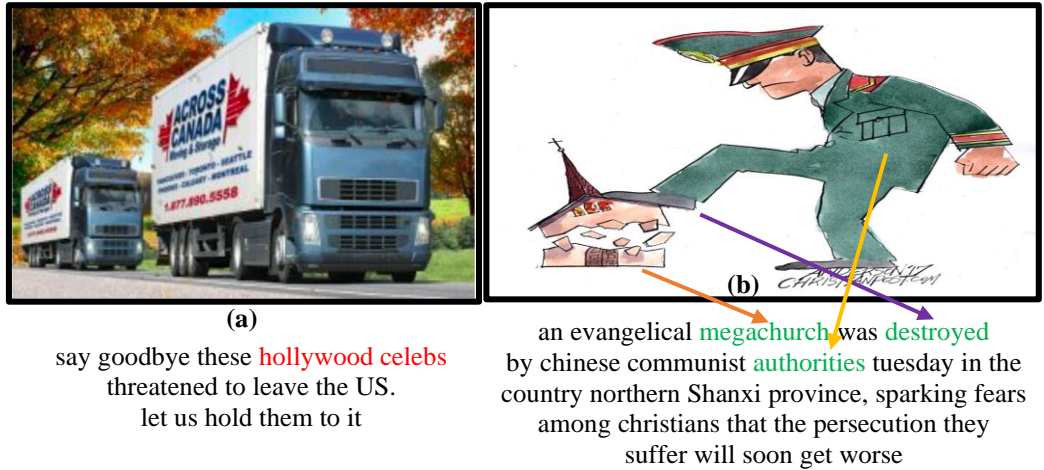

**Figure 1 Example of (a) Fake news sample (b) Real news sample from Pontes Fake News Dataset**

Figure 1 shows some sample image-text pairs of (a) Fake news and (b) Real news from Pontes Fake News Dataset [5]. Firstly, it is difficult to predict the authenticity of fake news just by looking at the image or reading the text description due to the heterogeneity of these modalities. It is necessary to train the model by combining the features from both modalities and learning unique characteristics from each of them. Secondly, due to the unstructured form of the language found on social media and the subjectivity in the human recognition process, it is not easy to focus on the essential words and the corresponding image regions in the samples. Hence, it is crucial to capture the correlation of the multimodal data, which provides clues for detecting fake news. As shown in Figure 1 (a), the textual part contains words like *hollywood, celebs* which are nowhere seen in the corresponding image. On the other hand, in Figure 1 (b), the words like *megachurch, destroyed, authorities* in the textual part are strongly related to specific regions shown in the corresponding images. Hence, more focus should be given to explore the semantic correlation between the images and their corresponding text.

Some of the prior works in multimodal fake news detection have used pre-trained CNN architectures like VGG-16, ResNet to extract the visual features, and LSTM, RNNs for capturing the textual features [6] [2]. This is followed by early-level fusion or feature-based fusion and late fusion-based methods [1]. However, they fail to capture the complementary information between the multiple modalities. Attention mechanisms have shown tremendous success in deep learning for tasks like visual question answering [7], object classification [8] [9], human motion prediction [10], etc. These networks are inspired by the visual perception of humans and can focus on the crucial regions by assigning them more weights, which helps to enhance the overall performance of the classifier [11]. Moreover, the latest transformer-based methods [12] are showing tremendous success in multimodal data classification.

In this paper, we propose an Efficient Transformer Based Multilevel Attention (ETMA) framework, that comprises three major components: visual-attention-based encoder, text-attention-based encoder, and joint attention-based learning. First, spatial information is added corresponding to the different patches of the input image. The visual-attention based encoder learns the abstract features from the embedded patches by applying multi-head self-attention, multi-layer perceptron, and layer normalization. In parallel, the text-attention based encoder generates a comprehensive embedding by combining the token embeddings, segment embeddings, and position embeddings, which generates the contextual information from the input text sequence. Finally, the visual semantic attention block models the correlations from the visual and textual data by selecting the unique image features based on the attended text features. The redundant features are further removed by using self-attention. The significant contributions of this paper are summarized as follows:

- To propose an Efficient Transformer Based Multilevel Attention framework (ETMA) for detecting fake news from the multimodal data by uniquely combining the power of attention mechanisms at different levels.
- To exploit the complementary information between the image and text modalities by extracting the image features based on the attended text-based features. The redundancy is removed by using self-attention, which eliminates the unnecessary features, thereby reducing the dimensionality of the data.
- Experiments are conducted on four-real world fake news datasets: Twitter, Jruvika Fake News Dataset, Pontes Fake News Dataset, and Risdal Fake News Dataset. The model was validated using multiple evaluation metrics like Accuracy, Precision, Recall, F1 scores, ROC curves, and Area under ROC curves. Experimental results prove the efficacy of the proposed model.
- An ablation study is performed to analyze the importance of different components in the proposed model. The computation time of the proposed model is also lower than the other baseline methods.

The rest of the manuscript is organized as follows: Section 2 discusses the related work in the area of fake news detection. Section 3 explains the proposed architecture. In section 4, we apply different evaluation metrics to validate the performance of our model. Finally, section 5 discusses the conclusion and future scope of the current work.

## 2. Related Works

This section explores the literature related to the fake news classification. Since deep learning-based methods offer promising solutions in this area, we majorly discuss the baseline methods related to deep-based unimodal and multimodal fake news detection.

## 2.1 Unimodal fake news detection

Jae-Seung Shim *et al.* [13] proposed a context-based approach that utilizes the network information of the user and vectorizes it by using link2vec strategy. The authenticity of the links generated by any web-based search was examined by studying their composition pattern. The link2vec approach was an extension to the existing word2vec approach for generating the embeddings of the words. Experiments on two language-based datasets for detecting fake news show that the link-based approach beats all other baseline methods. Lianwei Wu *et al.* [14] proposed a category-controlled encoder-decoder approach that produces differentiated samples based on the target category. The news samples guided the encoder and performed semantic matching to generate the semantic context representations of the data. The decoder captures the intra-category discriminative features. Thus, the module highlights the inter-category features and strengthens the intra-category features. The experimental results indicated that the proposed method significantly outperforms the previous methods.

Trueman *et al.* [15] proposed an attention-based methodology for detecting fake news on several datasets. The authors used convolution-based Bi-LSTM networks in conjunction with the attention-based method to capture the global, local, and temporal features from the data. However, the authors have used the traditional attention-based approaches, which fail to perform well with the data. Paka *et al.* [16] developed a COVID-19 dataset from Twitter and used semi-supervised methods inspired by attention networks to detect fake news. The proposed model uses the tweet-based and user-based features and embeds them with the external knowledge corresponding to each tweet. However, the model was tested only on the COVID-19 fake news data and failed to generalize well with the generic fake data found on Twitter.

Verma *et al.* [17] investigate the linguistic features that help in classifying the news as fake or real. Dong *et al.* [18] proposed a semi-supervised deep-based network that utilizes CNNs to extract the low-level features with a limited labeled dataset. The authors utilize cross-entropy loss and mean-squared loss to validate the proposed network. Liao *et al.* [19] developed a fabricated news recognition multitask learning model based on the perception that a few specific topics have higher rates of fake news, and the news creators have higher goals to distribute fake news. The proposed model considers the effect of subject names for fake news and presents context-oriented data to support the fast identification of fake news. In particular, the proposed model was inspired by representation learning and multitask learning for identifying fake news and classifying the news topics.

**2.2 Multimodal fake news detection**

Singhal *et al.* [2] developed SpotFake, a multimodal approach to detect the fake news from full-length articles. The authors applied a pre-trained VGG-16 model to extract the features from images and XLNet to extract the textual features. Xue *et al.* [20] proposed a Multimodal Consistency Neural Network (MCNN) that considers the consistency of multimodal information and catches the general attributes from web-based media data. The author developed a visual tampering module to extract the tampered features from the images. The MCNN computes the similarity between the modalities to extract the semantic and physical features.

Song *et al.* [21] utilized multichannel CNN and residual attention networks, which removes the noise from the multimodal data and enhances the performance of the classifier. The authors conducted experiments on Twitter and three variants of Weibo datasets. The accuracy of 74%, 85.3%, 86.9%, and 92.2% was achieved on Twitter, Weibo-A, Weibo-B, and Weibo-C datasets, respectively. Yuan *et al.* [22] improved the fake news detection task by learning the domain-invariant features of fake data generated on different events. The feature extractor fetches the multimodal features from the data, and the domain discriminator differentiates between the information spread across multiple domains. They further developed a relational graph structure, where the news was represented in the form of nodes, and the labels were written on the links joining the nodes of the graph.

Meel *et al.* [23] exploited the hidden structure of the textual data by applying hierarchical attention networks. The image captioning block looks into the headlines of the news articles, and the noise block removes the noise from the visual data. The ensemble-based approach utilizes max voting for determining the final output category. Kumari *et al.* [6] proposed a model which consists of different submodules: attention network based on Bi-directional LSTM extracts the textual features, Attention network based on CNN and RNN extracts the visual features, the bilinear pooling layer generates the multimodal features, and the multi-level perceptron for the final classification. The major drawback of the proposed method was that it was unable to perform well on longer text sequences and the model also failed to identify the semantic correlations between the multimodal features, resulting in misclassification errors.

Li *et al.* [1] proposed an entity-specific method that focuses on paying attention to specific objects from the multimodal data and aligning them using capsule units and dynamic routing algorithms. The method focuses on modeling the relationship between the entities and extracting the multimodal features, which helps identify the authenticity of the input. Jin *et al.* [24] developed an attention-based RNN method that fuses the image features extracted with VGG-16 model with the text features and social contextual data obtained using the LSTM network. Khattar *et al.* [25] proposed a multimodal variational autoencoder (MVAE) comprising the encoder and decoder modules. The encoder utilizes Bi-LSTMs to extract the textual information and VGG-16 to get the visual features and generates a latent vector, which was passed into the decoder module. The decoder later helps reconstruct the original data, and the fake news detector helps to predict whether the input is fake or real. Unlike these works, we propose a framework to model the semantic correlations between the image and text modalities to avoid any mismatch between the multimodal features. This would capture the invariant features from the complex images by keeping the distinctive characteristics of each modality.

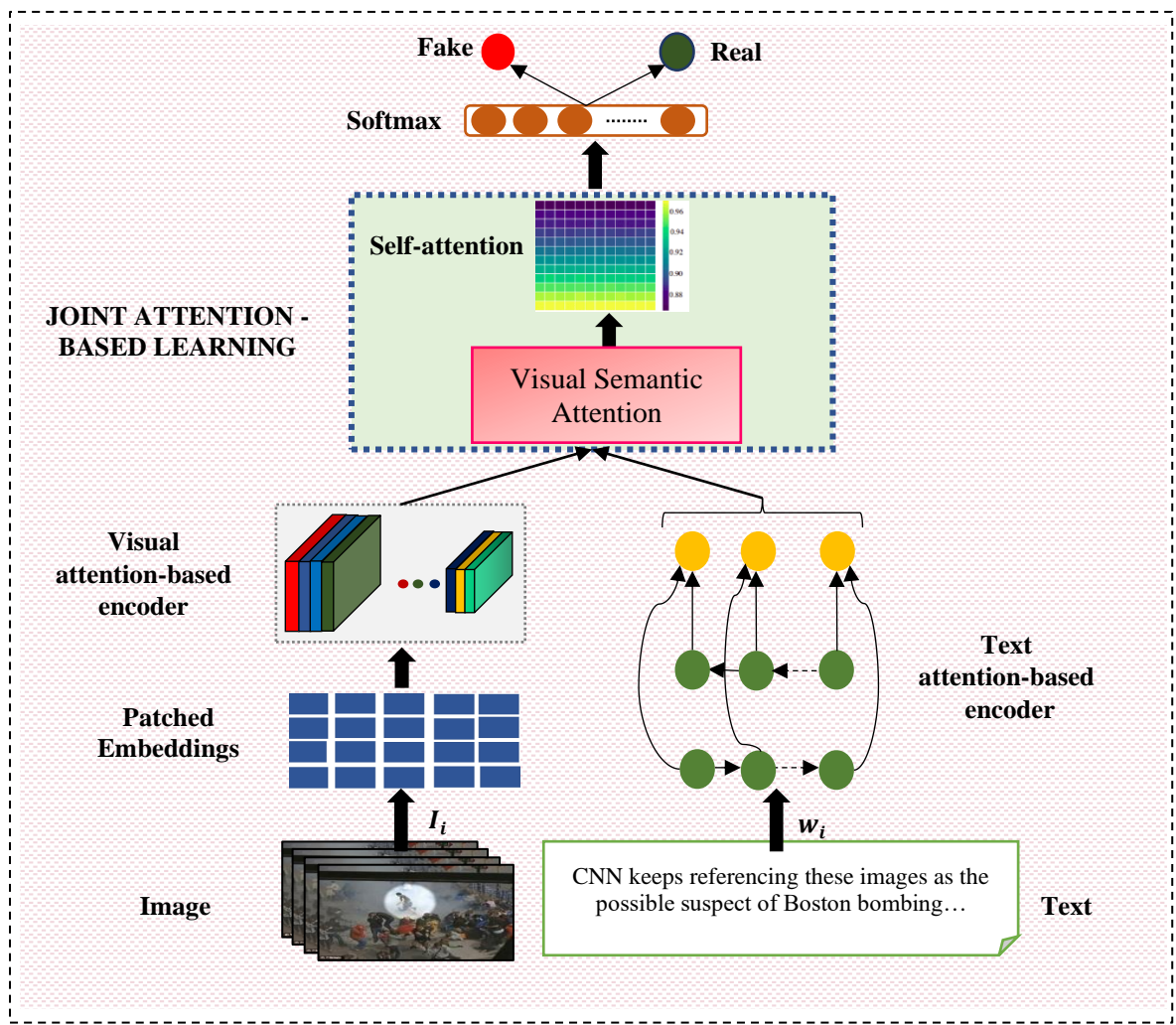

**Figure 2 Proposed Efficient Transformer Based Multilevel Attention (ETMA) framework**

## 3. PROPOSED METHODOLOGY

### 3.1 Task Definition

Given a set of multimodal samples $M = \{m_1, m_2, \ldots, m_n\}$. Each $m_i \in M$, contains sentences with $w_i$ words and an image $I_i$ with the associated target $T_i$. Each target $T_i$ is attached with a label $y_i$, which can be either real or fake. Initially, we removed those instances from the multimodal datasets that contain either text data or image data, so that we could have a uniform distribution of both the modalities. Text and images are preprocessed separately. Text data is preprocessed by using the NLTK library, which helps in removing the stop words and converting the word into its root form by using stemming and lemmatization. The longer sentences are divided into smaller tokens by applying tokenization. Images are normalized by performing mean subtraction and scaling. We have also used data augmentation techniques like rotation, flipping, zooming, etc., so that the model does not overfit the training data. Our task is to predict the correct label for the set of unseen samples with the help of the proposed ETMA framework, as shown in Figure 2.

### 3.2 Patch embeddings

Each image $I_i$ is divided into small patches, where each patch leverages a 16*16 convolution with a stride of 16. The batch of input images with shape $(b, h, w, c)$ is divided into fixed-size patches, which are flattened to generate the flat patches. We multiply these patches with a trainable embedding vector of dimension $d$. This gives us a low-dimensional linear embedding of the flattened patches. A learnable token is also prepended to the patch embeddings to get a combined representation of all the patches. Then we add the positional embeddings so that the transformer model has complete knowledge about the sequence of images. In this way, we add the spatial information corresponding to each patch in the sequence. The entire process is summarized in Figure 3:

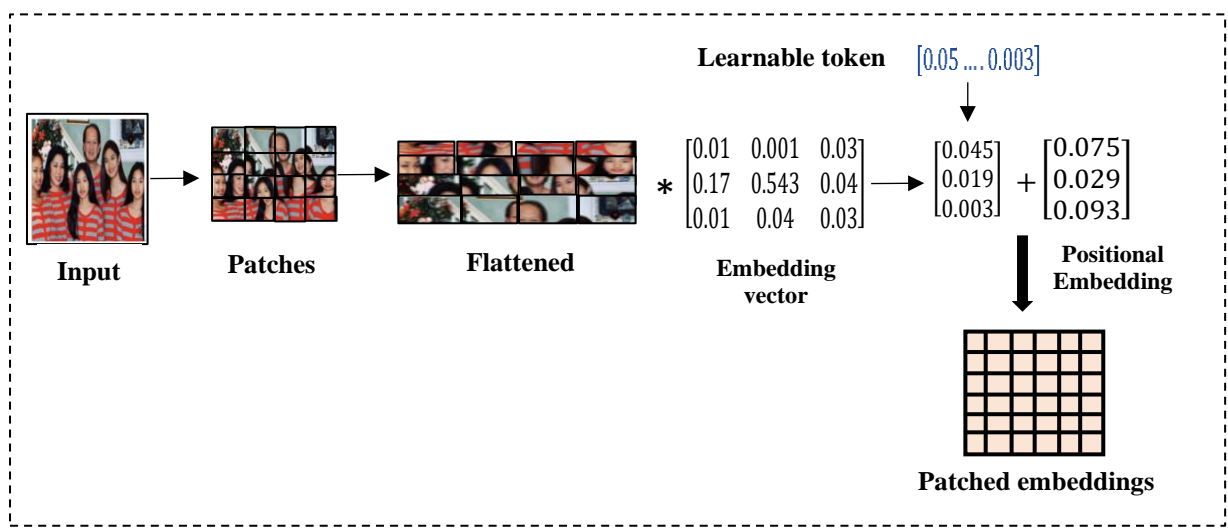

**Figure 3 Generating Patch embeddings from the visual data**

### 3.3 Visual attention-based encoder mechanism

The patched embeddings generated in Section 3.2 are passed into the transformer attention-based encoder, which learns the abstract features from the patches. For the visual data, we have used the *Vision transformer* [26] as the backbone architecture. The encoder module basically includes the following components [27]: Multi-head self-attention (MSA), multi-layer perceptron (MLP), and layer normalization (Norm). The benefit of self-attention is that it can capture the information globally from the entire image. So, the multi-head self-attention block divides the inputs into multiple heads, where each head can learn and understand the different aspects of the abstract representation of the input. We combine the output of all the heads and pass them into the MLP layer, which uses the GeLu non-linearity. Layer normalization is applied before every layer to reduce the training time of the network. Residual connections are also applied to avoid the vanishing gradient problem. The entire process is summarized as follows:

$$\acute{z}_l = MSA\left(Norm\left(\acute{z}_{l-1}\right)\right) + \acute{z}_{l-1} \qquad (1)$$

$$z_l = MLP\left(Norm\left(\acute{z}_l\right)\right) + \acute{z}_l \qquad \forall\, l = 1,2,\ldots L \qquad (2)$$

The $'+'$ operation denotes the residual connection. The embedded patches are denoted by $\acute{z}_{l-1}$. $z_l$ goes as input to the next encoder. The layer normalization operation includes scaling with mean and standard deviation for each input.

The MLP contains one hidden layer and one output layer. The mapping of input $X$ to output $O$ using MLP is shown in Eq (3) as follows:

$$O = GeLu\,(W_0\,X + b_0) \qquad (3)$$

The *MSA* is the multi-head self-attention which is structured as follows:

$$MSA = W_0\,[head_1, head_2, \ldots, head_n] \qquad (4)$$

and, $head_i = softmax\left(\frac{Q\cdot K}{\sqrt{d}}\right).V$

where $[head_1, head_2, \ldots, head_n]$ are the multiple heads. $Q, K, V$ are the trainable matrices and signify how the input can be projected into three sub-spaces, $d =$ dimension of each head.

### 3.4 Text attention-based encoder mechanism

The raw text sequences are encoded using the bidirectional encoder representation from Transformers [28], which again utilizes the attention mechanism. The text sequences are converted into tokens by combining the token embeddings, segment embeddings, and position embeddings. The token embeddings ($T_i$) give the vocabulary IDs of each token, the sentence embeddings ($S_i$) help in differentiating between two sentences, and the position embeddings ($P_i$) indicates the position of each word in the sentence. Each embedding layer contains different multi-head self-attention sublayers and is connected to the previous sublayers. Thus, for a sequence of $n$ words, the final embeddings $E_{f_i}$ can be summarized as follows:

$$E_{f_i} = \{T_i + S_i + P_i\} \; \forall \; i = 1, \ldots, n \tag{5}$$

The $E_{f_i}$ gives a comprehensive and robust embedding containing information about the input. The sequence of the hidden states can be defined as $H = \{h_1, h_2, \ldots, h_n\}$ and the output of the MSA is defined as:

$$\alpha_i(h) = Softmax\left(\frac{(w_q * h)(w_k * h)}{\sqrt{d/m}}\right)(w_v * h)^T \tag{6}$$

$w_q, w_k, w_v$ are model parameters corresponding to the query, key, and values, respectively and $m$ represents the number of attention mechanisms which are combined as follows:

$$H = [\alpha_i(h), \ldots, \alpha_m(h)]^T \tag{7}$$

The layer normalization and residual connections are added and stacked together. Finally, the output of the last layers gives the contextual information for the input sequence.

### 3.5 Joint attention-based learning

The existing works in multimodal analysis fail to learn the discriminative features between the image and text modalities. In multimodal feature learning, it becomes crucial to explore the complementary information between the different modalities. This will enhance the overall performance of our model. The joint attention-based learning uses two modules to achieve this. The first module is the visual semantic attention block which extracts the crucial image features based on attended text features to generate the multimodal features. The second module comprises of self-attention block, which removes the redundant features from the multimodal data. Each of them is explained in the subsequent sections.

### 3.5.1 Visual Semantic attention

Given an image-text pair $\{I_i, T_i\}$, for an $i^{th}$ multimodal sample, we pass them into the visual semantic attention block, which aims to learn the attention on the image features based on the words contained in the text sequence. This is done by fusing both the modalities with the help of element-wise multiplication as shown below:

$$vs_i = (\alpha^T I_i) * (\beta^T T_i) \tag{8}$$

$\alpha^T, \beta^T$ are the learnable parameters and $*$ denotes element-wise multiplication. The attention scores are calculated as follows:

$$\mathfrak{a}_i = \frac{\exp(\sigma(w^T vs_i + b))}{\sum_i \exp(\sigma(w^T vs_i + b))} \tag{9}$$

These attention scores signify the attentive strength on the different regions of the images. Finally, the attended image-level features are calculated as follows:

$$\mathcal{F}_i = \sum_i \mathfrak{a}_i * I_i \tag{10}$$

The $\mathcal{F}_i$ signifies the image-level multimodal features and signifies the image regions corresponding to the different words in the text sequence.

### 3.5.2 Self-attention

In the self-attention block, multiple modalities interact with each other to tell which feature should be given more importance and compute attention of all the inputs with respect to each other. This is very important since many irrelevant features might be generated while combining the modalities. The multimodal features (comprising image and text features) are allowed to interact with each other to find the features that need more importance. Hence, the self-attention block will highlight the different multimodal features according to their weights by combining the attention of all the inputs with respect to each other. Mathematically, this can be written as:

$$P_i = \frac{\exp(\varphi(W * \mathcal{F}_i + b))}{\sum_i (\rho(W * \mathcal{F}_i + b))} \tag{11}$$

The weighted average of all the self-attended multimodal features is calculated as:

$$M_i = \sum_i \mathcal{F}_i * P_i \tag{12}$$

The final obtained features are passed into the softmax classifier for the classification. Softmax is a probabilistic activation function. It gives the probability of the class membership for each output label. The output having the maximum probability is selected as the final class. The softmax function is expressed as follows:

$$Softmax\ (x_i) = \frac{\exp(x_i)}{\sum_j \exp(x_j)} \tag{13}$$

Here, $x$ = the value of neurons in the output layer. The values are divided by the sum of the exponential value, which normalizes and converts them into probabilities. The network is trained by minimizing the cross-entropy loss as shown in Eq (9):

$$Loss = -\sum_{j=1}^{2}(y_j \cdot \log \hat{y}_j) \tag{14}$$

Here, $y_j$ is the observed output corresponding to the actual output $\hat{y}_j$ for the $j^{th}$ sample.

## 4. Experiments

This section discusses the experimental results and baseline comparison of the proposed ETMA framework with four-real world datasets for multimodal fake news classification.

### 4.1 Implementation Details

The complete architecture is implemented in Python language on 64-bit Windows 10 machine with 128 GB RAM NVIDIA Titan RTX GPUs. The hyperparameter details are shown in Table 1:

**Table 1 Hyperparameter Details of the proposed ETMA**

| Hyperparameter | Twitter Dataset [29] | Jruvika Fake News Dataset [30] | Pontes Fake News Dataset [5] | Risdal Fake News Dataset [31] |
|---|---|---|---|---|
| Image size | 224*224*3 | 224*224*3 | 224*224*3 | 224*224*3 |
| Learning rate | 0.001 | 0.001 | 0.003 | 0.0005 |
| Batch size | 128 | 64 | 128 | 128 |
| #epochs | 120 | 80 | 100 | 100 |
| Optimizer | Adam | Adam | Adam | Adam |
| Dropout | 0.5 | 0.3 | 0.5 | 0.4 |
| Joint-loss function | Cross entropy | Cross entropy | Cross entropy | Cross entropy |

### 4.2 Datasets

In order to show the efficacy of our proposed architecture, we have conducted experiments on four real-world datasets, namely Twitter, Jruvika Fake News Dataset, Pontes Fake News Dataset, and Risdal Fake News Dataset. The statistics of each dataset are summarized in Table 2 as follows:

**Table 2 Statistics of multimodal fake news datasets**

| Dataset | # of real samples | # of fake samples | Total |
|---|---|---|---|
| Twitter [29] | 5,910 | 7,420 | 13,330 |
| Jruvika Fake News Dataset [30] | 1,861 | 2,012 | 3,873 |
| Pontes Fake News Dataset [5] | 25,249 | 20,136 | 45,385 |
| Risdal Fake News Dataset [31] | 8,071 | 11,834 | 19,905 |

### 4.2.1 Twitter

This dataset is released in the "Verifying multimedia use" [31] task for detecting the fake content on the Twitter platform. The dataset is divided into training and testing set based on different events. The dataset has 7,420 fake tweets and 5,910 real tweets. We use 10% of the training set as the validation set and consider only those samples that contain both the text and its associated image.

### 4.2.2 Jruvika Fake News Dataset

This dataset is downloaded from Kaggle [30] and contains URL, headline, body, and output label for the various news articles. The dataset doesn't contain the images, so we use the URL to download the corresponding images using Python's Beautiful Soup library. The headline and body are combined together and passed into the text module of our proposed system. The dataset contains 2,012 fake news and 1,861 real news articles.

### 4.2.3 Pontes Fake News Dataset

Guilherme Pontes dataset [5] is also hosted on Kaggle and contains news articles that are divided into several categories like clickbait, fake, rumor, reliable, etc. We have categorized the fake, unreliable, clickbait, rumor into the Fake news category and reliable into the real news category. The corresponding images are downloaded from the URLs associated with each news instance. We only keep those instances that have both image and text pairs. This gives us 25,249 real news and 20,136 fake news instances.

### 4.2.4 Risdal Fake News Dataset

Megan Risdal's Fake News dataset [31] is also downloaded from Kaggle and contains 20,000 news articles with 11,834 fake news instances and 8,071 real news instances. The dataset contains the URL of the images, which we download using the Python library.

### 4.3 Baseline Comparison

In this section, we compare our work with the following baselines to validate the effectiveness of our proposed model. The results are shown in Table 3 to Table 6.

**Table 3 Comparative analysis of the ETMA with baseline methods on Twitter dataset (Acc: Accuracy, P: Precision, R: Recall, F1: F1 score)**

| Datasets | Models | Acc | Real News | | | Fake News | | |
|---|---|---|---|---|---|---|---|---|
| | | | P | R | F1 | P | R | F1 |
| Twitter | VQA [32] | 0.68 | 0.60 | 0.77 | 0.67 | 0.78 | 0.61 | 0.68 |
| | EMAF [1] | 0.80 | 0.73 | 0.86 | 0.79 | 0.88 | 0.75 | 0.81 |
| | Att-RNN [24] | 0.74 | 0.67 | 0.88 | 0.76 | 0.85 | 0.61 | 0.71 |
| | MVAE [25] | 0.74 | 0.68 | 0.77 | 0.73 | 0.80 | 0.71 | 0.75 |
| | Spotfake [2]: | 0.88 | 0.87 | 0.76 | 0.81 | 0.89 | 0.95 | 0.92 |
| | DAGA-NN [22] | 0.90 | 0.91 | 0.77 | 0.83 | 0.89 | 0.96 | 0.93 |
| | Ensemble with Max Voting [23] | 0.88 | 0.89 | 0.90 | 0.91 | 0.90 | 0.87 | 0.89 |
| | MCNN [20] | 0.89 | 0.92 | 0.90 | 0.91 | 0.89 | 0.85 | 0.87 |
| | AMFB [6] | 0.74 | 0.67 | 0.88 | 0.76 | 0.85 | 0.61 | 0.71 |
| | CARM-N [21] | 0.91 | 0.87 | 0.91 | 0.89 | 0.89 | 0.95 | 0.92 |
| | **Ours** | **0.93** | **0.93** | **0.92** | **0.92** | **0.94** | **0.92** | **0.93** |

**Table 4 Comparative analysis of the ETMA with baseline methods on Jruvika Fake News dataset (Acc: Accuracy, P: Precision, R: Recall, F1: F1 score)**

| | Models | Acc | P | R | F1 | P | R | F1 |
|---|---|---|---|---|---|---|---|---|
| Jruvika Fake News Dataset | VQA [32] | 0.62 | 0.66 | 0.69 | 0.67 | 0.61 | 0.67 | 0.64 |
| | EMAF [1] | 0.77 | 0.70 | 0.73 | 0.71 | 0.76 | 0.80 | 0.78 |
| | Att-RNN [24] | 0.72 | 0.62 | 0.65 | 0.63 | 0.63 | 0.68 | 0.65 |
| | MVAE [25] | 0.79 | 0.75 | 0.83 | 0.79 | 0.82 | 0.84 | 0.83 |
| | Spotfake [2]: | 0.77 | 0.81 | 0.79 | 0.80 | 0.84 | 0.83 | 0.83 |
| | DAGA-NN [22] | 0.83 | 0.84 | 0.88 | 0.86 | 0.87 | 0.89 | 0.88 |
| | Ensemble with Max Voting [23] | 0.92 | 0.93 | 0.89 | 0.91 | 0.92 | 0.91 | 0.91 |
| | MCNN [20] | 0.92 | 0.95 | 0.93 | 0.94 | 0.89 | 0.92 | 0.90 |
| | AMFB [6] | 0.85 | 0.84 | 0.83 | 0.83 | 0.86 | 0.85 | 0.85 |
| | CARM-N [21] | 0.94 | 0.95 | 0.93 | 0.94 | 0.91 | 0.92 | 0.91 |
| | **Ours** | **0.97** | **0.96** | **0.97** | **0.96** | **0.95** | **0.94** | **0.94** |

**Table 5 Comparative analysis of the ETMA with baseline methods on Pontes Fake News Dataset (Acc: Accuracy, P: Precision, R: Recall, F1: F1 score)**

| | Models | Acc | P | R | F1 | P | R | F1 |
|---|---|---|---|---|---|---|---|---|
| Pontes Fake News Dataset | VQA [32] | 0.67 | 0.65 | 0.70 | 0.67 | 0.64 | 0.66 | 0.65 |
| | EMAF [1] | 0.71 | 0.68 | 0.69 | 0.68 | 0.65 | 0.67 | 0.66 |
| | Att-RNN [24] | 0.74 | 0.75 | 0.77 | 0.76 | 0.71 | 0.74 | 0.72 |
| | MVAE [25] | 0.77 | 0.74 | 0.72 | 0.73 | 0.79 | 0.81 | 0.80 |
| | Spotfake [2]: | 0.81 | 0.77 | 0.72 | 0.74 | 0.75 | 0.74 | 0.74 |
| | DAGA-NN [22] | 0.83 | 0.81 | 0.83 | 0.82 | 0.79 | 0.80 | 0.79 |
| | Ensemble with Max Voting [23] | 0.94 | 0.93 | 0.94 | 0.93 | 0.95 | 0.91 | 0.93 |
| | MCNN [20] | 0.91 | 0.89 | 0.86 | 0.87 | 0.91 | 0.92 | 0.91 |
| | AMFB [6] | 0.87 | 0.89 | 0.91 | 0.90 | 0.93 | 0.90 | 0.91 |
| | CARM-N [21] | 0.95 | 0.92 | 0.87 | 0.89 | 0.91 | 0.94 | 0.92 |
| | **Ours** | **0.96** | **0.95** | **0.94** | **0.94** | **0.95** | **0.93** | **0.94** |

**Table 6 Comparative analysis of the ETMA with baseline methods on Risdal Fake News Dataset (Acc: Accuracy, P: Precision, R: Recall, F1: F1 score)**

| | | Acc | P | R | F1 | P | R | F1 |
|---|---|---|---|---|---|---|---|---|
| Risdal Fake News Dataset | VQA [32] | 0.79 | 0.78 | 0.81 | 0.79 | 0.82 | 0.78 | 0.80 |
| | EMAF [1] | 0.81 | 0.77 | 0.79 | 0.78 | 0.84 | 0.83 | 0.83 |
| | Att-RNN [24] | 0.83 | 0.81 | 0.80 | 0.80 | 0.79 | 0.76 | 0.77 |
| | MVAE [25] | 0.76 | 0.75 | 0.77 | 0.76 | 0.73 | 0.76 | 0.74 |
| | Spotfake [2]: | 0.84 | 0.86 | 0.82 | 0.84 | 0.80 | 0.83 | 0.81 |
| | DAGA-NN [22] | 0.83 | 0.81 | 0.83 | 0.82 | 0.79 | 0.81 | 0.80 |
| | Ensemble with Max Voting [23] | 0.89 | 0.87 | 0.86 | 0.86 | 0.89 | 0.87 | 0.88 |
| | MCNN [20] | 0.91 | 0.88 | 0.87 | 0.87 | 0.85 | 0.89 | 0.87 |
| | AMFB [6] | 0.89 | 0.88 | 0.85 | 0.86 | 0.83 | 0.87 | 0.85 |
| | CARM-N [21] | 0.92 | 0.89 | 0.90 | 0.89 | 0.85 | 0.90 | 0.87 |
| | **Ours** | **0.95** | **0.97** | **0.96** | **0.96** | **0.98** | **0.97** | **0.97** |

- **VQA** [32]: In the visual-question answering task, the answers are provided corresponding to a given image. Since VQA is a multi-class problem, we replace the multi-class classifier with the binary classifier.
- **EMAF** [1]: The authors consider the entity-centric interactions among the multimodal data. The semantic entities are aligned with the visual entities using the dynamic routing algorithm.
- **Att-RNN** [24]: The image features extracted from the pre-trained CNN model were fused with the text-based features and social context information obtained using the attention mechanism and LSTMs network.
- **MVAE** [25]: Variational auto-encoder is used to encode the information inside the textual and visual data using BiLSTMs and VGG-19 networks, respectively.
- **Spotfake** [2]: The authors used pre-trained XLNet to get the text feature vector and VGG-19 network for extracting the visual patterns.
- **DAGA-NN** [22]: Graph-attention neural network is developed to learn the domain invariant features of fake news. The image features are extracted by a pre-trained VGG-19 network, and the BERT model learns the textual features.
- **Ensemble with Max Voting** [23]: The ensemble of text and visual features are used to extract the fake content from the input. The Hierarchical attention network extracts the text-based features, and an image caption generator creates the image summary.
- **MCNN** [20]**:** The authors measure the similarity of the multimodal data by evaluating the correlation between the multimodal features.
- **AMFB** [6]: The textual features are extracted by the attention-based BiLSTM, which are stacked and combined together. The visual features are extracted by the multi-channel CNN-RNN unit, incorporating the attention mechanism. However, the authors did not consider semantic attention for longer sentences.

- **CARM-N** [21]: The authors applied a cross-attention residual network to extract the features from the data and multi-channel CNN to remove unnecessary noise generated by fusing the multimodal features.

The results clearly show that our proposed model outperforms the state-of-the-art and other baseline methods. We observe that the lowest performance is given by baseline [32]. This might be because the authors did not apply attention mechanisms, so the model could not focus on the crucial regions during classification. For our proposed model, we observe that the highest accuracy is obtained on the Jruvika Fake News dataset, which is 3% more than the accuracy of [23]. Our model also gives the best performance in terms of precision, recall, and F1 scores evaluation metrics for all other datasets. This clearly shows that our model can keep the unique characteristics of each modality, which is very important when we deal with multimodal datasets.

**4.4 Results and analysis**

Figure 4 shows the classification results of the proposed method on the different fake news datasets. We report the model's accuracy along with the precision, recall, and F1 scores corresponding to each of the output classes. Our model achieves an accuracy of 93%, 97%, 96%, and 95% on Twitter, Jruvika fake news detection, Pontes fake news dataset, and Risdal fake news dataset, respectively. This verifies the robustness of our model as the visual semantic attention block can capture the complex correlation patterns between the image and text. This avoids any mismatch between the features of the multimodal data, and the fake data can be captured effectively.

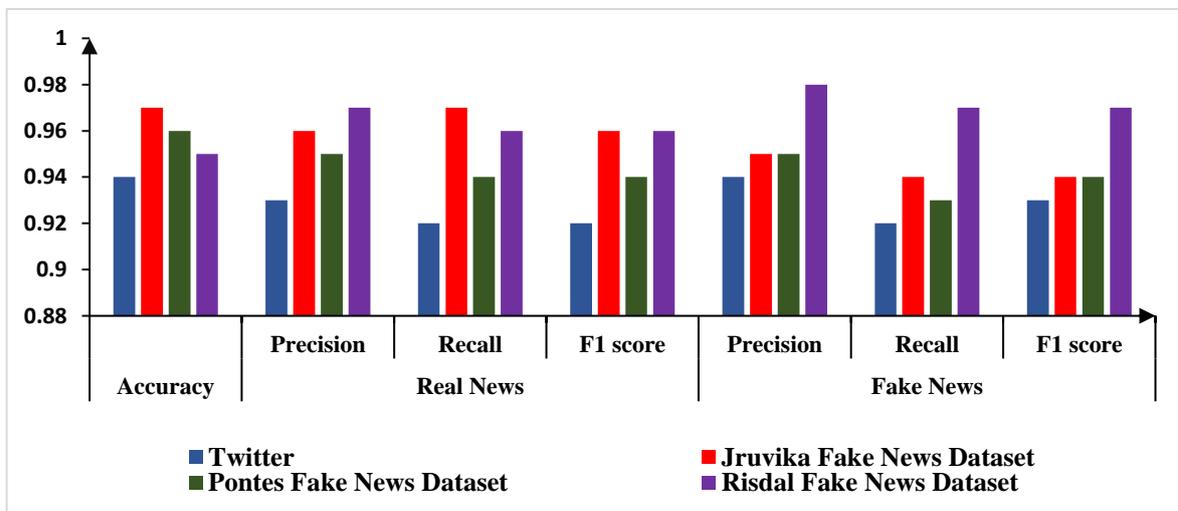

**Figure 4 Classification results on the Datasets**

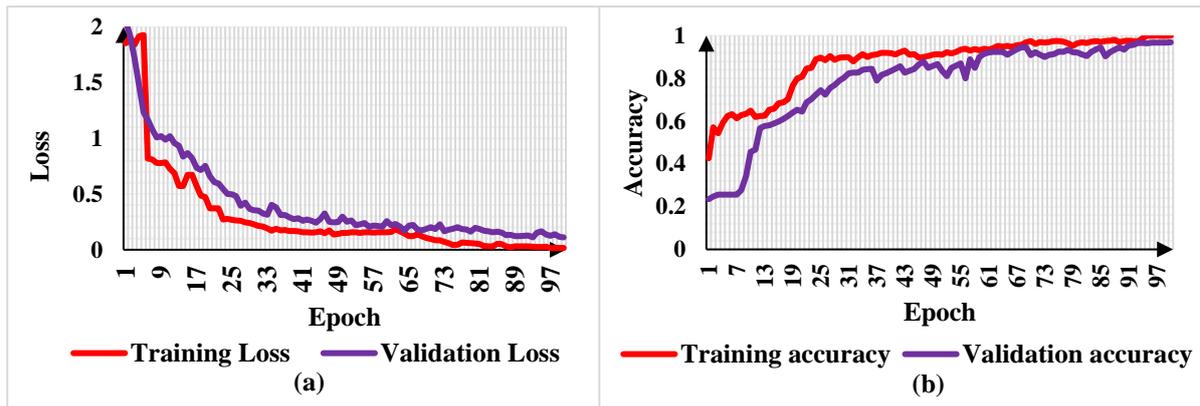

**Figure 5 Training and Validation (a) Loss curves (b) Accuracy curves for our model**

The learning curves for our model are shown in Figure 5. The training and validation loss curves, shown in Figure 5 (a), decrease with the epochs, and the training and validation accuracy, in Figure 5 (b), increase with the number of epochs. The training accuracy is 100%, and the training loss is 0%, which means our model is trained to its full capacity from the supplied data, and the performance is validated by monitoring the validation accuracy and validation loss. This confirms the adequate learning of our model. The learning curves signify that as we supply more data to the model, it can learn more features and eventually converge in 100 epochs. The validation accuracy for each dataset is monitored continuously, and the model achieving the highest accuracy is selected as the final model for the test dataset. The receiver output characteristic (ROC) curve, shown in Figure 6, is plotted between the True Positive Rate and the False Positive Rate (FPR) values. These curves are essential as they also reflect the classifier's true performance with the unbalanced data samples. The area under the curve (AUC) metric helps to compare the different ROC curves. We observe that the AUC values are 0.98, 0.94, 0.93, and 0.92 for Jruvika fake news dataset, Pontes Fake news dataset, Risdal fake news dataset, and Twitter dataset, respectively. This confirms that the model is not affected by the unbalanced samples in any of the datasets. ROC curves can sometimes give misleading results in case of imbalanced datasets. Hence, we again evaluated the performance of our classifier by using the Precision-Recall (PR) curves which is a stricter metric to validate the performance of the classifier. These curves are specially used to see whether the model is performing well in case of imbalanced dataset by plotting the Precision and Recall values and comparing the Area under the curve (AUC) value. As seen in Figure 7, we have plotted the PR curve for the Pontes Fake News dataset as in this dataset, the proportion of real news and fake news samples is the largest. The AUC=0.94 shows that the model is not affected by the imbalanced data samples in the dataset.

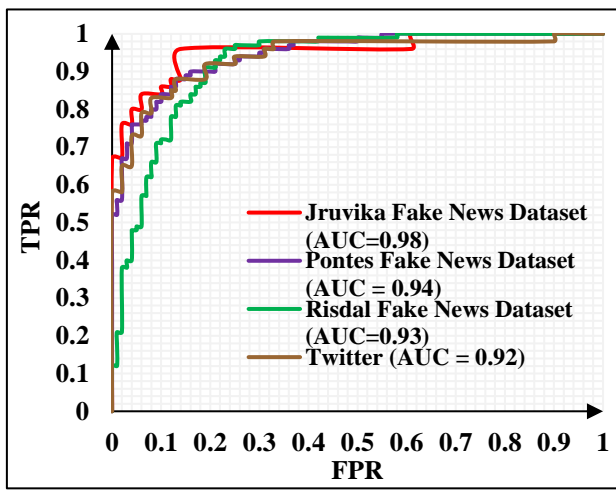

**Figure 6 ROC curves for the datasets**

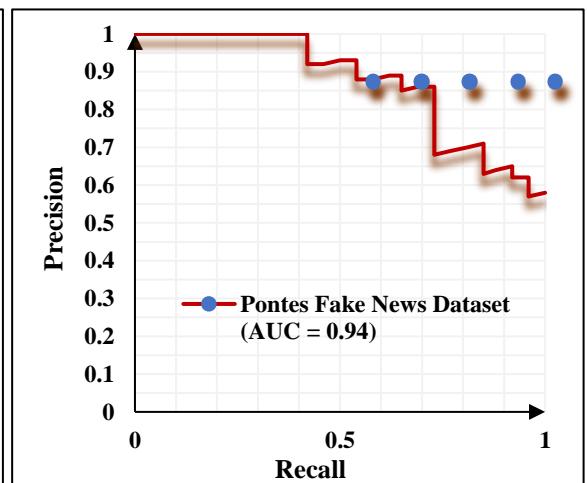

**Figure 7 Precision Recall (PR) Curve for Pontes Fake News Dataset**

Our proposed model can capture the invariant features from the complex images, and the uniqueness of each modality is intact even after the fusion. Moreover, the self-attention block removes the redundant features, thus drastically enhancing our model's performance.

**4.5 Ablation study**

In this section, we conduct an ablation study to analyze the effect of different components in our proposed model. We first conduct the unimodal analysis on textual and visual data separately and then perform the multimodal analysis on all the datasets. The results are summarized in Table 7.

- **Unimodal analysis:** The input is passed into a text attention-based encoder, followed by self-attention (Self-attn) for the textual modality. The extracted features are passed into the softmax layer for the final classification. In the case of visual modality, we generate the patched embeddings and pass them to the visual attention-based encoder module, followed by the self-attention mechanism. The final features are directly passed into the softmax classifier. In both these cases, the visual semantic attention (VS-attn) block is ablated as we are dealing with unimodal data only.
- **Multimodal analysis:** In multimodal analysis, we conduct experiments by ablating different components in our proposed architecture to analyze the importance of each component. We first remove the self-attention block from the architecture and directly pass the multimodal features from the visual semantic block to the softmax classifier. Then we remove the visual-semantic attention block from the architecture. We observe that in this case, maximum accuracy is dropped. The accuracy achieved on Twitter, Jruvika Fake News Dataset, Pontes Fake News Dataset, and Risdal Fake News Dataset is 80%, 85%, 90%, and 87%, respectively. This shows the importance of incorporating the semantic correlation between the image and text features. Then we remove the visual attention-based encoder block, pass the patched embeddings through the pre-trained VGG-16 model, and then perform the joint-attention-based learning. The results clearly validate the importance of our visual attention-based encoder block, which captures the invariant features from the images. Finally, we ablate the text attention-based encoder and see that text-based attention is significant when dealing with sentences as the crucial words are highlighted, which helps to establish the context.

**Table 7 Ablation study of the proposed architecture**

| Modality | Model | Accuracy | | | |
|---|---|---|---|---|---|
| | | Twitter | Jruvika Fake News Dataset | Pontes Fake News Dataset | Risdal Fake News Dataset |
| Unimodal | Textual | 0.79 | 0.81 | 0.79 | 0.84 |
| | Visual | 0.83 | 0.83 | 0.87 | 0.86 |
| Multimodal | Model w/o Self-attn | 0.90 | 0.95 | 0.94 | 0.93 |
| | Model w/o VS-attn | 0.80 | 0.85 | 0.90 | 0.87 |
| | Model w/o Visual attention-based encoder | 0.86 | 0.87 | 0.92 | 0.89 |
| | Model w/o Text attention-based encoder | 0.88 | 0.92 | 0.93 | 0.91 |
| | **Ours** | **0.93** | **0.97** | **0.96** | **0.95** |

## 4.6 Computation time

We verify the performance of the proposed model in real-time, which includes generating the multimodal feature vector and the testing time for each of the multimodal input samples. The experiments were conducted on a 64-bit Windows 10 machine with 128 GB RAM and NVIDIA Titan-RTX GPUs. The results are summarised in Table 8. We compare our results with the CARM-N [21], which has shown the highest results amongst all the baseline methods (Refer Table 3 to Table 6). Our proposed methods take 11.6 ms to extract the multimodal features and 0.46 ms for testing the input sample. These results show that our proposed method is faster in terms of generating the features and testing the sample to detect fake news from multimodal data. Hence, our proposed method has shown proficient performance in terms of all evaluation metrics and is also time-efficient compared to other baseline methods.

**Table 8 Computation time analysis of the proposed architecture**

| Method | Feature formulation per sample (Text + image) | Testing time per sample (Text + image) |
|---|---|---|
| CARM-N [21] | 35.8 ms | 0.74 ms |
| Ours | 11.6 ms | 0.46 ms |

### 4.7 Case study

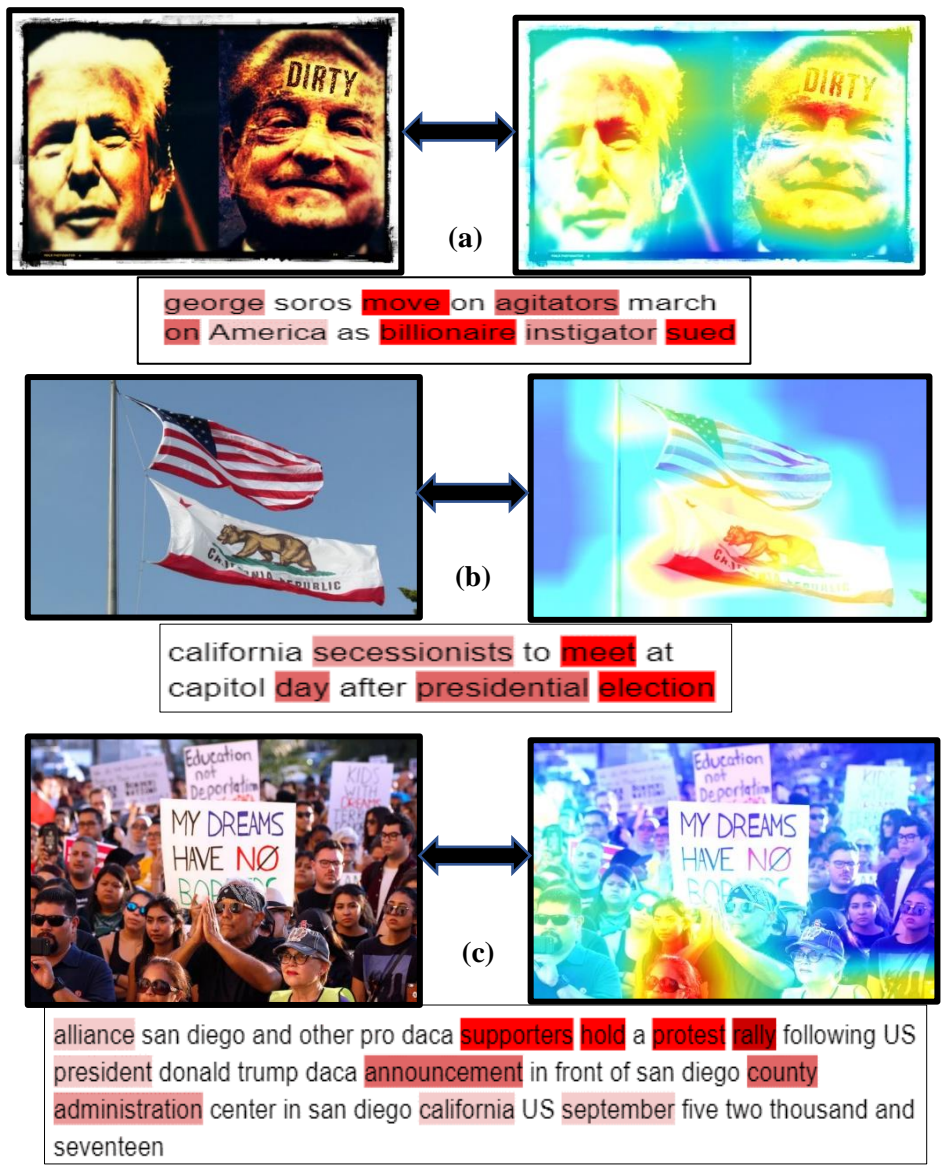

**Figure 8 (a-b) Visualizing the attention in (a) and (b) Fake samples and (c) Real samples**

We show the importance of applying attention mechanisms at multiple levels in the proposed architecture by visualizing the weights of visual-semantic attention block. In figure 8, we visualize the regions in the images and the words in the sentences, which are getting more importance and help in identifying the fake news. In Figure 8 (a) and (b), we observe that the words do not correlate with the content shown in the images. Words like *agitators* are not getting mapped with the input image in (a), and words like *election, meet* are not correlated with the content of the input image in (b). Thus, the model classifies them as fake samples. On the other hand, in (c), words like *supporters, protest, rally* are more focused by the model as they are directly linked with the image content. Thus, we observe that visual semantic attention block in the proposed model can relate the crucial words to the image regions, enhancing our model's classification ability.

## 5. Conclusion and Future Scope

In this work, we explored the problem of learning complementary information between the multimodal data by proposing an Efficient Transformer based Multilevel Attention framework (ETMA). The network utilizes attention mechanisms at multiple levels and focuses on crucial regions in the images based on the attended textual features. We also employ self-attention at the end to remove any redundancy from the multimodal data. The experimental results conducted on four popular real-world datasets show that our method performs efficiently, and the model has less computation time as compared to state-of-the-art methods.

The multimodal fake news classification area is still unexplored and requires more attention from researchers. Hence, in the future, a more comprehensive fusion technique can be developed that incorporates the social network information along with the image-text pairs. Moreover, future researchers can explore event-based multimodal fake news detection where the multimodal pairs do not have a strong correlation among them.

## References

[1] P. Li, X. Sun, H. Yu, Y. Tian and F. Yao, "Entity-Oriented Multi-Modal Alignment and Fusion Network for Fake News Detection," *IEEE Transactions on Multimedia,* vol. 14, no. 8, pp. 1-14, 2015.

[2] S. Singhal, A. Kabra, M. Sharma, R. R. Shah, T. Chakraborty and P. Kumaraguru, "SpotFake+: A Multimodal Framework for Fake News Detection via Transfer Learning (Student Abstract)," *Proceedings of the AAAI Conference on Artificial Intelligence,* vol. 34, no. 10, pp. 13915-13916, 2020.

[3] L. Ying, H. Yu, J. Wang, J. Yongze and S. Qian, "Fake News Detection via Multi-Modal Topic Memory Network," *IEEE Access,* vol. 9, pp. 132818-132829, 2021.

[4] S. Hakak, M. Alazab, S. Khan, T. R. Gadekallu, P. Maddikunta and W. Khan, "An ensemble machine learning approach through effective feature extraction to classify fake news," *Future Generation Computer Systems,* vol. 117, pp. 47-58, 2021.

[5] "Pontes Fake News Dataset," [Online]. Available: https://www.kaggle.com/pontes/fake-news-sample.

[6] R. Kumari and A. Ekbal, "AMFB: Attention based multimodal Factorized Bilinear Pooling for multimodal Fake News Detection," *Expert Systems With Applications,* vol. 184, pp. 1-12, 2021.

[7] J. Yu, W. Zhang, Y. Lu, Z. Qin and Y. Hu, "Reasoning on the Relation: Enhancing Visual Representation for Visual Question Answering and Cross-Modal Retrieval," *IEEE Transactions on Multimedia,* vol. 22, no. 12, pp. 3196-3209, 2020.

[8] Y. Zhu, C. Zhao, H. Guo and J. Wang, "Attention CoupleNet: Fully Convolutional Attention Coupling Network for Object Detection," *IEEE Transactions on Image Processing,* vol. 28, no. 1, pp. 113-126, 2019.

[9] D. Zhao, Y. Chen and L. Lv, "Deep Reinforcement Learning With Visual Attention for Vehicle Classification," *IEEE Transactions on Cognitive and Developemental Systems,* vol. 9, no. 4, pp. 356-367, 2017.

[10] X. Chen and W. Wang, "Uni-and-Bi-Directional Video Prediction via Learning Object-Centric Transformation," *IEEE Transactions on Multimedia,* vol. 22, no. 6, pp. 1591-1604, 2020.

[11] A. Yadav and D. K. Vishwakarma, "A Language-independent Network to Analyze the Impact of COVID-19 on the World via Sentiment Analysis," *ACM Transactions on Internet Technology,* vol. 22, no. 1, pp. 1-30, 2021.

[12] L. Zheng, B. Liu and J. Tao, "CTNet: Conversational Transformer Network for Emotion Recognition," *IEEE/ACM Transactions on Audio, Speech, and Language Processing,* vol. 29, pp. 985-1000, 2021.

[13] J.-S. Shim, Y. Lee and H. Ahn, "A link2vec-based fake news detection model using web search results," *Expert Systems With Applications,* vol. 184, pp. 1-15, 2021.

[14] L. Wu, Y. Rao, C. Zhang, Y. Zhao and A. Nazir, "Category-controlled Encoder-Decoder for Fake News Detection," *IEEE Transactions on Knowledge and Data Engineering,* pp. 1-14, 2021.

[15] T. E. Trueman, A. Kumar, N. P. and V. J., "Attention-based C-BiLSTM for fake news detection," *Applied Soft Computing,* vol. 110, pp. 1-8, 2021.

[16] W. Paka, R. Bansal, A. Kaushik, S. Sengupta and T. Chakraborty, "Cross-SEAN: A cross-stitch semi-supervised neural attention model for COVID-19 fake news detection," *Applied Soft Computing,* vol. 107, pp. 1-13, 2021.

[17] P. Verma, P. Agrawal, I. Amorim and R. Prodan, "WELFake: Word Embedding Over Linguistic Features for Fake News Detection," *IEEE Transactions on Computational Social Systems,* vol. 8, no. 4, pp. 1-13, 2021.

[18] X. Dong, U. Victor and L. Qian, "Two-Path Deep Semisupervised Learning for Timely Fake News Detection," *IEEE Transactions on Computational Social Systems,* vol. 7, no. 6, pp. 1386-1398, 2020.

[19] Q. Liao, H. Chai, H. Han, X. Zhang and X. Wang, "An Integrated Multi-Task Model for Fake News Detection," *IEEE Transactions on Knowledge and Data Engineering,* pp. 1-12, 2021.

[20] J. Xue, Y. Wang, Y. Tian, Y. Li, L. Shi and L. Wei, "Detecting fake news by exploring the consistency of multimodal data," *Information Processing and Management,* vol. 58, no. 5, pp. 1-13, 2021.

[21] C. Song, N. Ning, Y. Zhang and B. Wu, "A Multimodal Fake News Detection Model Based on Crossmodal Attention," *Information Processing & Management,* vol. 58, no. 1, pp. 1-35, 2021.

[22] H. Yuan, J. Zheng, Q. Ye, Y. Qian and Y. Zhang, "Improving fake news detection with domain-adversarial and graph-attention neural network," *Decision Support Systems,* vol. 151, pp. 1-11, 2021.

[23] P. Meel and D. K. Vishwakarma, "HAN, image captioning, and forensics ensemble multimodal fake news detection," *Information Sciences,* vol. 567, pp. 23-41, 2021.

[24] Z. Jin, J. Cao, H. Guo, J. Luo and Y. Zhang, "Multimodal Fusion with Recurrent Neural Networks for Rumor Detection on Microblogs," *Proceedings of the 25th ACM international conference on Multimedia.,* pp. 795-816, 2017.

[25] D. Khattar, J. Goud, M. Gupta and V. Varma, "MVAE: Multimodal Variational Autoencoder for Fake News detection," *The world wide web conference,* pp. 2915-2921, 2019.

[26] A. Dosovitskiy, L. Beyer, A. Kolesnikov, D. Weissenborn, X. Zhai, T. Unterthiner, M. Dehghani, M. Minderer, G. Heigold, S. Gelly, J. Uszkoreit and N. Houlsby, "An image is worth 16x16 words: Transformers for image recognition at scale," *Ninth International Conference on Learning Representations (ICLR),* pp. 1-22, 2021.

[27] A. Dosovitskiy, L. Beyer, A. Kolesnikov, D. Weissenborn and X. Zhai, "An image is worth 16x16 words: Transformers for image recognition at scale.," *Proceedings of the International Conference on Learning Representations,* pp. 1-22, 2021.

[28] J. Devlin, M.-W. Chang, K. Lee and K. Toutanova, "BERT: Pre-training of Deep Bidirectional Transformers for Language Understanding," *Annual Conference of the North American Chapter of the Association for Computational Linguistics,* pp. 1-16, 2019.

[29] C. Boididou, S. Papadopoulos, D. Dang-Nguyen, G. Boato and M. Riegler, "Verifying multimedia use at mediaeval," *Working Notes Proceedings of the MediaEval 2016,* pp. 1-3, 2016.

[30] "Jruvika Fake News Dataset," [Online]. Available: https://www.kaggle.com/jruvika/fake-news-detection.

[31] Y. Yang, L. Zheng, J. Zhang, Q. Cui and X. Zhang, "TI-CNN: convolutional neural networks for fake news detection," *arXiv preprint arXiv:1806.00749,* pp. 1-11, 2018.

[32] A. Agrawal, J. Lu, S. Antol, M. Mitchell, L. Zitnick, D. Batra and D. Parikh, "VQA: Visual Question Answering," *Proceedings of the IEEE international conference on computer vision,* pp. 2425-2433, 2015.